\documentclass{IEEEtran}

\usepackage{etoolbox}
\usepackage{amsmath}
\usepackage{amssymb}
\usepackage{booktabs}
\usepackage{multirow}
\usepackage{siunitx}
\usepackage{graphicx}
\usepackage{array}
\usepackage{caption}
\usepackage{subcaption}
\usepackage{setspace}
\usepackage{balance}
\usepackage{environ}
\usepackage{multicol}
\usepackage{algorithm}
\usepackage{algpseudocode}
\usepackage{url}
\usepackage{cite}
\usepackage[english]{babel}

\usepackage{booktabs}
\usepackage[table]{xcolor}
\usepackage{rotating} 

\NewEnviron{IEEEimpact}{%
  \section*{IMPACT STATEMENT}
  \BODY
}

\newcommand{\tightesttabular}{\setlength{\tabcolsep}{1.5pt}\scriptsize}

\title{\textbf{\textsc{Tempus}}: A \textbf{Temp}orally \textbf{S}calable Resource-Invariant GEMM \textbf{S}treaming Framework for Versal AI Edge}

\author{\IEEEauthorblockN{Mahdieh Grailoo$^{*}$}
\IEEEauthorblockN{José Núñez-Yáñez$^{\dagger}$}\\
\IEEEauthorblockA{$^{*}$\textit{Linköping University}, Sweden\\
$^{\dagger}$\textit{Universidad Politecnica de Madrid}, Spain\\
$^{*}$mahdieh.grailoo@liu.se, $^{\dagger}$jose.nunez.yanez@upm.es}}

\begin{document}
\maketitle

\begin{abstract}
Scaling laws for Large Language Models (LLMs) establish that model quality improves with computational scale, yet edge deployment imposes strict constraints on compute, memory, and power. Since General Matrix Multiplication (GEMM) accounts for up to $90\%$ of inference time, efficient GEMM acceleration is critical for edge AI. The Adaptive Intelligent Engines available in the AMD Versal adaptive SoCs are well suited for this task, but existing state-of-the-art (SOTA) frameworks maximize performance through spatial scaling, distributing workloads across hundreds of cores — an approach that fails on resource-limited edge SoCs due to physical implementation failures, bandwidth saturation, and excessive resource consumption.
We propose \textsc{Tempus}, a Resource-Invariant Temporal GEMM framework for the AMD Versal AI Edge SoC. Rather than expanding hardware resources with matrix size, \textsc{Tempus} employs a fixed compute block of 16 AIE-ML cores, achieving scalability through iterative graph execution and algorithmic data tiling and replication in the Programmable Logic. High-speed cascade streaming ensures low-latency partial sum reduction at Initiation Interval (II) of 1, while a deadlock-free DATAFLOW protocol maximizes transfer-compute overlap and PLIO reuse.
Evaluated on GEMM workloads, \textsc{Tempus} achieves $607$ GOPS at $10.677$ W total on-chip power. By characterizing system-level efficiency through the Platform-Aware Utility (PAU) metric, we prove that \textsc{Tempus} achieves a $211.2\times$ higher prominence factor than the leading spatial SOTA (\textsc{ARIES}). Furthermore, the framework maintains a $0.00\%$ utilization of URAM/DSP, yielding $22.0\times$ core frugality, $7.1\times$ power frugality, and a $6.3\times$ reduction in I/O demand, establishing a sustainable, scalable foundation for edge LLM inference.
\end{abstract}

\begin{IEEEkeywords}
Hardware-Software Co-design, Versal ACAP System-on-Chip, Intelligent Engine, Large Language Models, Temporal Scaling, Resource-Constrained Edge Devices, Parallelization, Sustainable Inference.
\end{IEEEkeywords}

\section{INTRODUCTION} \label{sec:introduction}
The scaling laws of Large Language Models (LLMs) have demonstrated that even with access to a large resource pool, temporal scaling is the only viable path for complete model deployments \cite{hoffmann2022training,kaplan2020scaling,pearce2024reconciling}. 
Therefore the unprecedented scale and computational demands of modern LLMs, require specialized hardware acceleration for deployments \cite{xu2025slim,jiang2025comprehensive,guo2025survey,li2025tiny,nunez2025sgrace,grailoo2024heterogeneous}, particularly in the constrained edge devices \cite{CHARMtwo,ARIES,AutoMM,CHARM,AutoSA}. In these models, the efficiency of $\text{General Matrix Multiplication}$ ($\text{GEMM}$) is the central performance bottleneck of workloads, typically consuming over $90\%$ of the total execution time during inference. \textsc{Tempus} focuses on sustainable acceleration of rectangular GEMM on AMD Versal Adaptive Compute Acceleration ($\text{ACAP}$) Edge platforms which are heterogeneous System-on-chips consisting of a processing system (PS), programmable logic (PL) and adaptable intelligent engines (AIE-ML)\cite{AMD_Versal_AI_Edge_PSG_2024,AMD_UG1079_2025,Xilinx_Versal_AI_Edge_WP518_2021}.

Prior state-of-the-art (SOTA) optimization frameworks designed for Versal ACAPs competed on achieving peak throughput by relying on massive spatial scaling, distributing the workload across hundreds of intelligent engines, typically found on larger devices like VC Versal Core and VE Versal, which host 300 to 400 cores \cite{aie4ml,ARIES,AutoMM,AutoSA,CHARM,CHARMtwo,GAMA,MaxEVA}. This approach fundamentally fails when ported to resource-limited edge devices. These designs require high core and resource utilization, leading to excessive power consumption and saturation of scarce PL components.
This saturation confines the use of the PL fabric for integrating essential non-MM kernels (like Softmax or Layer normalization), needed for complete model inference \cite{grailoo2024activation}. Furthermore, pushing spatial limits often leads to physical implementation failures. 
Therefore, the inherent assumption that performance scales linearly with core count breaks down in this constrained context.

To overcome the performance/resource mismatch at the edge, we introduce Temporal rectangular GEMM Scaling, a novel framework that achieves high performance and scalability by limiting and fixing hardware resource allocation, prioritizing efficiency over spatial parallelism.
Our major contributions are as follows.
\begin{enumerate}

\item \text{Resource-Invariant Frugality Framework:} \textsc{Tempus} decouples resource utilization from matrix size by considering a fixed spatial compute block. The scaling for large workloads is achieved via iterative AIE-ML graph execution and algorithmic data replication. In addition, the 3D MatMul structure maps onto a fixed 2D array (e.g., Split$\times$Cascade) using data reduction and multi-casting. Versus SOTA, \textsc{Tempus} achieves core, power, and I/O frugality. Also, by restricting programmable logic to lightweight streaming FIFOs and fixed-size tiling buffers, the architecture uses $0.00\%$ of URAM/DSP, preserving fabric for non-GEMM kernels (e.g., Softmax, Layer-normalization) required by foundation models.

\item Platform-Aware Utility and Architectural Efficiency: \textsc{Tempus} prioritizes architectural proficiency by normalizing performance against physical potential via the Platform-Aware Utility (PAU) metric, achieving a $211.2\times$ higher prominence than the leading spatial SOTA. Evaluation demonstrates near-ideal scaling, where a $32,\!768\times$ workload increase results in only a $6.8\times$ latency growth, effectively amortizing fixed system initialization costs. We further show that efficiency is modulated by the micro-kernel dimension (DIM), with optimized tile sizes yielding a $10.5\times$ latency reduction; a figure that can be further improved with additional local memory.

\item Compute-Transfer Overlapping Efficiency: High-speed streaming via a cascade interface enables low-latency partial sum reduction, avoiding $\sim$50\% slower buffer-sharing methods while guaranteeing a pipeline initiation interval (II) of 1. To circumvent the edge ``Bandwidth Wall,'' we maximize PLIO reuse through hybrid packet/broadcast switching and a deadlock-free DATAFLOW protocol. PL streaming additionally overlaps computation with data transfer between programmable logic and AIE arrays, effectively hiding communication latency.

\item Analytical Modeling of Performance-Critical Parameters: Our work introduces analytical models that derive the parameters, governing system-level efficiency. These models determine scheduling parameters, such as \texttt{GRAPH\_ITER\_CNT} for temporal scaling, Kernel size for tiling, and the Replication Factor for data reuse.

\end{enumerate}
The source code for this framework is openly available at \url{https://github.com/mgrailoo/TEMPUS}.

\begin{figure*}[t]
    \centering
    \includegraphics[width=\textwidth]{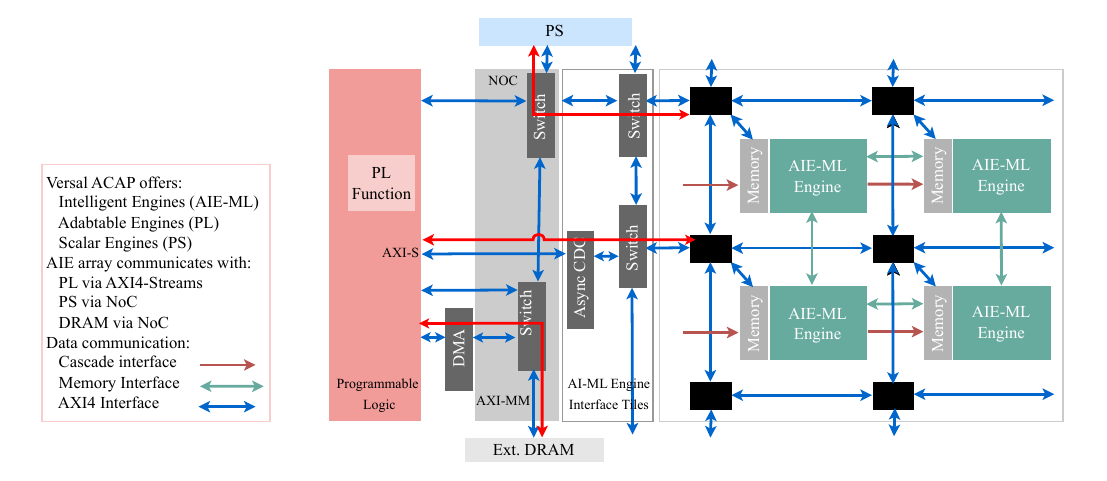}
\caption{Versal ACAP Architecture: Heterogeneous System Integration and Execution Flow for our framework}
    \label{fig:speedup_scaling}
\end{figure*}

\section{RELATED WORK: SPATIAL VS. TEMPORAL SCALING}
Prior GEMM acceleration on Versal ACAP evolved across two AI Engine generations, targeting maximized throughput via spatial scaling on large devices (300-400 cores). This philosophy fails on resource-limited edge platforms. Our work proposes temporal GEMM scaling as an alternative, resource-invariant paradigm.

\subsection{Spatial Scaling Frameworks and Utilization Challenges (Gen 1: AIE)}
These frameworks focused on achieving maximal theoretical performance on large AIE arrays, prioritizing throughput over resource frugality.

\begin{itemize}
    \item \textsc{CHARM} \& 2.0 (Heterogeneity-Aware Partitioning): \textsc{CHARM} pioneered GEMM on the AIE array using the Cascade Stream interface. The monolithic \textsc{CHARM} design suffered severe inefficiency with diverse layer sizes, suffering performance drops up to $5760\times$. \textsc{CHARM} 2.0 addressed this by partitioning the array into heterogeneous accelerators, improving BERT throughput by up to $5.29\times$. Using 288 AIE cores (72\% of VCK1902), 91.52\% BRAM and 82.94\% URAM Utilization, it achieved 10.03 TOPS on a $1024^3$ INT16 GEMM \cite{CHARM,CHARMtwo}. \textsc{CHARM} also faced significant resource issues, leading to certain INT8 designs utilizing only $48\%$ of AIE cores due to congestion problems \cite{CHARM}. 
    
    \item \textsc{MaxEVA} (Throughput-Centric Optimization): \textsc{MaxEVA} addressed the small matrix bottleneck, encountered by prior solutions like \textsc{CHARM}, and achieved high AIE-only throughput in simulation. However, it suffered limitations of using inefficient buffer-sharing, and dedicated AIE cores to reduction kernels, capping real-world efficiency. The pursuit of maximum spatial utilization led to physical implementation failure. For example, \textsc{MaxEVA}'s initial highest-throughput design, which required $100\%$ utilization of all 400 AIE cores, failed during Place-and-Route (PnR) due to routing congestion \cite{MaxEVA}.
    This simulation-focused approach provided theoretical performance, but lacked system implementation \cite{MaxEVA}.

    \item \textsc{AMA} (Algorithmic Efficiency): \textsc{AMA} is an advanced successor to \textsc{MaxEVA}, that eliminated dedicated reduction kernels by augmenting MAC kernels to perform accumulation directly. This innovation yielded performance and energy efficiency gain. However, \textsc{AMA} maintained the same fundamental limitation: it relied on the slower buffer-sharing interface for reduction, and its AIE-only simulation approach isolated it from real-world constraints despite using up to 342 cores \cite{AMA}.

    \item \textsc{AutoMM} (Resource-Conscious DSE): \textsc{AutoMM} introduced a resource-conservative design space exploration (DSE) for INT8/INT16 precision optimization, built on \textsc{CHARM}'s methodology. Utilizing 288 AIE cores (72\% of VCK1902), it achieved 7.51 TOPS with 56.8 W total power and lower BRAM utilization (49.33\%) versus spatial alternatives. However, its conservative resource approach capped performance scalability, with \textsc{ARIES} later demonstrating 1.57$\times$ superior energy efficiency for INT16 \cite{AutoMM}.
\end{itemize}

\subsection{Advanced Frameworks (Gen 2: AIE-ML) and Compiler-Aided Scaling}
As the architectural optimization space grew, compilation flows provided automated solutions to manage complex resource utilization patterns.

\begin{itemize}
    \item \textsc{GAMA} (AIE-ML Optimization): \textsc{GAMA} is the first study on second-generation of intelligent engines (AIE-ML) architecture \cite{GAMA}, i.e. VE2802. Its innovation was a custom buffer placement algorithm that achieved up to 100\% memory utilization, reducing stalls by 12\% versus standard compilers. Using staggered kernel placement to mitigate congestion, it achieved array utilization and performance in simulation. Critically, \textsc{GAMA} employed the faster Cascade interface, achieving higher throughput efficiency than \textsc{MaxEVA} and \textsc{ARIES}.
    
    \item \textsc{ARIES} (MLIR Compilation Flow): Introduced an agile MLIR-based flow for multi-level parallelism across Versal platforms \cite{ARIES}. Its core innovation was a unified MLIR representation spanning AIE and PL, enabling optimization and portability across AIE devices. Unlike simulation-based approaches, \textsc{ARIES} provided real on-board evaluation results. It achieved high throughput through massive spatial scaling, utilizing 88\% (352 cores) of AIEs with high PL resource usage (76\% URAM), making it unsuitable for resource-constrained edge devices. 
    \item \textsc{AutoSA} (Polyhedral Compilation): A polyhedral compiler generating monolithic systolic arrays with hardware optimizations (SIMD, II=1, double buffering) \cite{AutoSA}. While achieving high performance on $16nm$ AMD U250 FPGA, \textsc{CHARM} outperformed it with 2.9× higher throughput for the same precision.
\end{itemize}

In contrast to SOTA, our resource-invariant temporal scaling delivers performance through iterative execution, dimension reduction, and data replication. It is within a small, fixed core block using high-speed cascade and DATAFLOW streaming, ensuring resource conservation and edge compatibility.

\section{VERSAL ACAP ARCHITECTURE: The AI Edge VE2302}

The AMD Versal AI Edge VE2302 SoC integrates three distinct processing engines into a single heterogeneous architecture, as illustrated in Figure \ref{fig:speedup_scaling}. The Intelligent Engines form a 34-core array of VLIW/SIMD processors (AIE-ML), each with local memory \cite{lei2024mapping} (green-gray boxes), optimized for the deep learning compute kernels at the core of this architecture. The Adaptive Engines (Programmable Logic) (red box) provide the reconfigurable hardware (328K system logic cells, 464 DSPs), utilized for flexible logic and data movement, such as data streaming control (FIFOs), data tiling, de-tiling and replications. The Scalar Engines (Processing System) (blue box) incorporate dual-core Arm\textregistered~Cortex-A72 and Cortex-R5F processors for system orchestration and general-purpose tasks \cite{AMD_Versal_AI_Edge_PSG_2024,AMD_UG1079_2025,Xilinx_Versal_AI_Edge_WP518_2021}.

The AIE-ML array interfaces with the broader system through two key paths. It connects directly to the PL via high-speed AXI4-Streams (PLIO), which is the primary conduit for feeding data into the array. Communication with the Processing System (PS) and access to external DRAM are both facilitated through the high-bandwidth Network-on-Chip (NoC).
Within the AIE-ML array itself, three specialized data communication mechanisms enable efficient computation, which are fundamental to our scaling methodology. The Cascade Interface provides direct, low-latency connections (512-bit wide in AIE-ML) between adjacent cores  for rapid partial sum reduction, facilitating our temporal scaling approach. The Memory Interface enables buffer sharing between neighboring cores, while the AXI4 Switch connects non-adjacent cores and is configured for efficient packet-switching and broadcasting. For the sake of simplicity, all subsequent explanation, and diagrams will consider a 2x2 AIE-ML core array.

\section{METHODOLOGY: RESOURCE-INVARIANT TEMPORAL GEMM SCALING}
Our framework transforms large matrix multiplication into a predictable, iterative streaming process. By mapping the 3D MatMul ($\mathrm{GEMM\_SIZE\_A} \times \mathrm{GEMM\_SIZE\_AB} \times \mathrm{GEMM\_SIZE\_B}$) onto a fixed 2D AIE-ML array (Split$\times$Cascade, e.g., 2$\times$8) and employing a constant set of PL resources exclusively for dataflow, we achieve Resource-Invariant performance.

\subsection{System Orchestration and Control Flow (PS Side)}
The coordination of the heterogeneous Versal ACAP and the dataflow for our Temporal Scaling framework is managed by the Processing System/Host CPU, which acts as the central orchestrator as shown in Figure \ref{fig:speedup_scaling}. 
The process begins when the scalar engines initiate execution. Input matrices are stored in external DRAM. A dedicated DMA HLS kernel then manages data transfer from external DRAM to the AIE-ML array using a deadlock‑free DATAFLOW design, ensuring a continuous, high‑speed data stream. Data is transferred from external DRAM over the high‑bandwidth NoC (blue arrows) using the AXI4‑MM protocol. From there, data is streamed into the AIE‑ML array via the AXI4‑Stream (AXIS) network (red arrows).
Afterwards, data is streamed into the AIE-ML array via the AXI4-Stream (AXIS) network (red path). In this step, to maximize the reuse of scarce PLIO resources, specialized routing is employed (i.e., Broadcast circuit-switching and Packet Switching). While the fixed-core AIE-ML graph performs the matrix multiplication, internal AIE-to-AIE communication for partial sum reduction is handled by the high-speed, 512-bit Cascade Stream (dark red arrows in Figure \ref{fig:speedup_scaling}) \cite{AMD_UG1079_2025}. And the Memory Interface (green arrows) is used for local data access within each core's memory. Finally, the resulting Matrix C streams back from the AIE-ML array through the PL, where the DMA kernel collects it and writes it back to External DRAM via the NoC (red path), completing the execution cycle.

The detailed execution of the system is governed by a 7-phase timed control flow, orchestrated by the PS, in Algorithm~\ref{alg:host_execution}. In Phase~0 (INIT), the host calculates the critical \texttt{GRAPH\_ITER\_CNT} parameter for temporal scaling. It allocates memory buffers for matrices A, B, and C using XRT's aligned allocator, ensuring 4096-byte boundary alignment for optimal DMA performance. 
Phase~1-2 (DATA\_PREP, DEVICE\_INIT), load the generated PLIO streams and the hardware binary (.xclbin) onto the Versal device, initializing the AIE-ML array and PL kernel. Phase~3 (BUFFER\_CREATE) maps the allocated buffers to External DRAM, establishing host-device data channels. Phase~4 (DATA\_XFER\_HOST2DEV) transfers input matrices from host memory to device DRAM, instantiating the PL kernel and AIE graph.
The core computation begins with Phase~5 (KERNEL\_LAUNCH), launching the kernel, followed by Phase~6 (CORE\_COMPUTATION) where temporal scaling is enacted through iterative AIE graph execution (\texttt{gemm\_aie\_gr.run(GRAPH\_ITER\_CNT)}), concurrent with PL kernel operation (dma\_krnl.wait())). Finally, the host synchronizes completion,  and transfers results back to host memory.

\begin{algorithm}[!t]
\caption{Host Application Execution Flow (7 Phases)}
\label{alg:host_execution}
\begin{algorithmic}[1]
\Function{main}{argc, argv}
    \State PHASE 0: Configuration and Memory Setup
    \State Calculate \texttt{GRAPH\_ITER\_CNT} for temporal scaling.
    \State Load matrix data or generate test patterns.
    \State Allocate host memory using aligned\_allocator (4096-byte aligned).
    \State PHASE 1–4: Initialization and Setup
    \State Initialize XRT device and load XCLBIN.
    \State Create buffer objects and map them.
    \State Transfer A and B to device.
    \State Instantiate FPGA HLS kernel and AIE graph.
    \State PHASE 5: Kernel Launch \Comment{Start timer}
    \State Launch dma\_hls\_rhdl.
    \State PHASE 6: CORE COMPUTATION
    \State Run AIE graph for \texttt{GRAPH\_ITER\_CNT} times.
    \State Wait for kernel.
    \State Record compute total. \Comment{Stop timer}
    \State PHASE 7: Output/Validation 
    \State Sync output.
    \State Write output and validate.
    \State Print summary.
\EndFunction
\end{algorithmic}
\end{algorithm}

\subsection{Algorithmic Data Preparation (Tiling, Data Decomposition, and 3D-to-2D Mapping)}
\begin{figure*}[ht]
    \centering
    \includegraphics[width=1.0\textwidth]{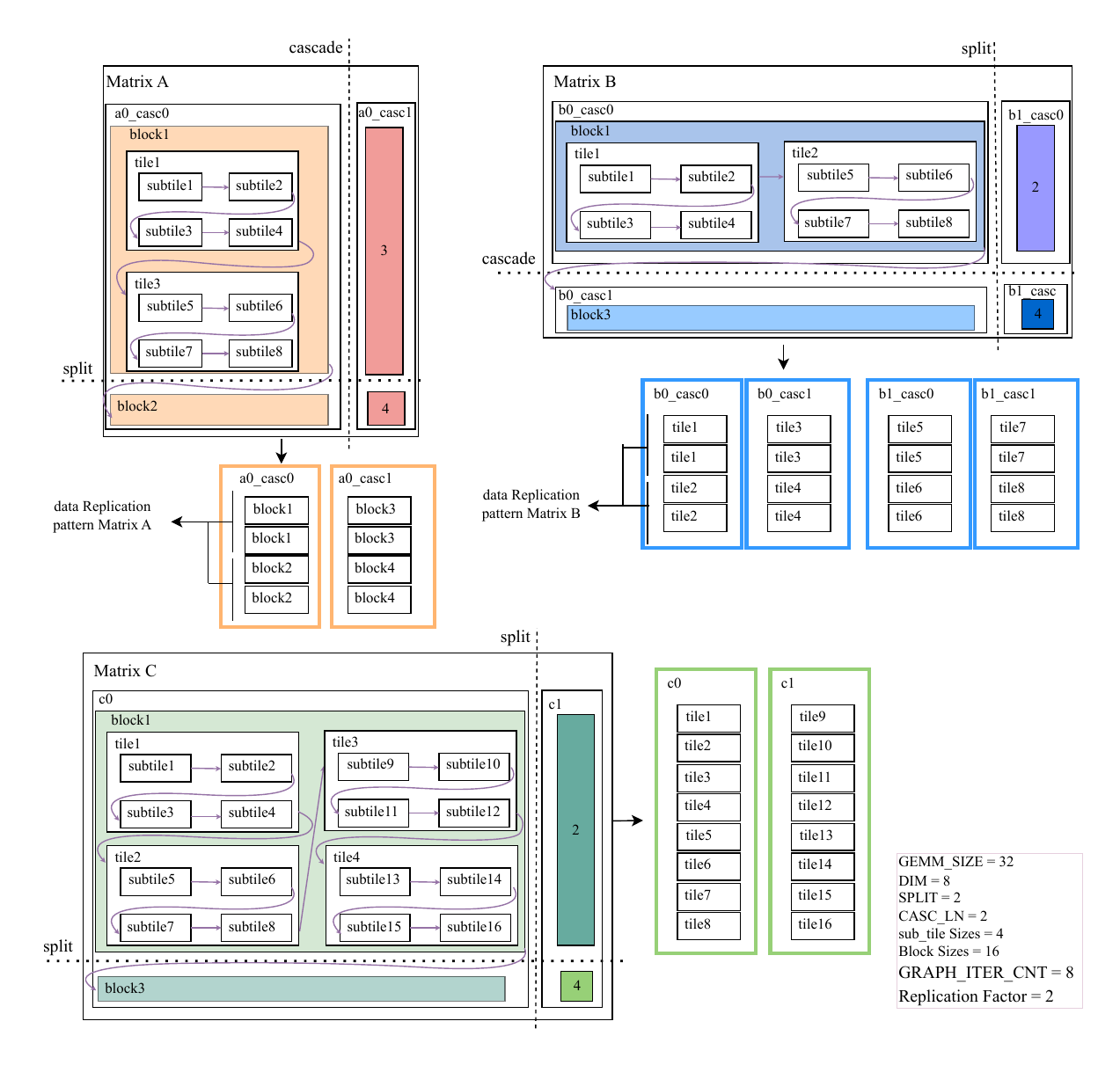}
    \caption{Hierarchical Data Decomposition and Stream Generation.}
    \label{fig:hierarchical_tiling}
\end{figure*}

This phase outlines how the framework overcomes hardware limitations, through dimension reduction, precise tiling, and specialized data repetition, as shown in Figure~\ref{fig:hierarchical_tiling}. 
The dimensional reduction maps the 3D MatMul ($\mathrm{GEMM\_SIZE\_A} \times \mathrm{GEMM\_SIZE\_AB} \times \mathrm{GEMM\_SIZE\_B}$) computation onto a fixed 2D core array ($\mathrm{SPLIT} \times \mathrm{CASC\_LN}$ cores), where $\mathrm{CASC\_LN}$ chains cores for $\mathrm{GEMM\_SIZE\_AB}$-dimension reduction via cascade streams, and $\mathrm{SPLIT}$ defines parallel groups. The $\mathrm{GEMM\_SIZE\_AB}$ dimension is processed through temporal iteration, while the $\mathrm{GEMM\_SIZE\_A}$ and $\mathrm{GEMM\_SIZE\_B}$ dimensions are distributed spatially across the array.

Figure~\ref{fig:hierarchical_tiling} illustrates the hierarchical decomposition strategy. The split boundaries (dotted lines) represent horizontal division into parallel processing groups for temporal scaling, while the cascade paths denote vertical AIE-to-AIE communication (governed by CASC\_LN) for partial sum reduction. The diagram shows large input matrices (A and B) converted into low-latency streams (a0\_casc0, a0\_casc1, b0\_casc0, b0\_casc1, b1\_casc0, b1\_casc1) organized hierarchically into Blocks (temporal units), Tiles (memory-defined by DIM), and Sub-tiles (vector units).
At the block level, matrices decompose into sequential Blocks (e.g., 'block1', 'block2'). Within blocks, data organize into Tiles ('tile1' through 'tile4') or micro-kernel corresponding to the DIM parameter. 
The maximum DIM is constrained by the AIE-ML core's local memory capacity, partitioned between matrices.

The smallest units are Subtiles ('subtile1' through 'subtile8'), representing minimal data segments optimized for AIE-ML vector execution. For the sake of simplicity, all subsequent diagrams, and explanation will consider a minimal $2\times2$ AIE-ML core array, rectangular GEMM size of $32\times16\times32$, DIM of 8, sub\_tile size of 4, and block size of $16\time8$, $8\time16$, and $16\time16$ for $A$, $B$ and $C$, respectively. During PLIO\_Cascade\_Stream\_Generation, Matrix A tiles follow row-major ordering while Matrix B uses column-major ordering, aligning with the core cluster's communication pattern. Matrix A replication occurs between tiles after each block processing cycle and Matrix B replication after each column within blocks. Subtile dimensions optimize for AIE-ML instruction efficiency, with physical dimensions adapting to DATA\_TYPE (e.g., 4$\times$4$\times$4 for int16 and int32), while maintaining row-major element serialization within subtiles.

\begin{algorithm}
\caption{PLIO Stream Generation and Tiling}
\label{alg:plio_generation}
\begin{algorithmic}[1]
\Require MatA, MatB; Config parameters (GEMM\_SIZE, DIM, SPLIT, CASC\_LN, DATA\_TYPE)
\Ensure Cascade input streams (a0\_casc*),  (b*\_casc*)
\State WRD\_LN $\gets 128 / \text{DATA\_TYPE}\_{\text{bits}}$ 
\Comment{Elements per 128-bit PLIO chunk}
\State \texttt{GRAPH\_ITER\_CNT} $\gets$ calculated via Equ. \ref{equ:gic}
\State Replication Factor $\gets$ calculated via Equ. \ref{equ:rf}
\For{each temporal block}
    \State Matrix A Ordering: Process MatA tiles in row-major order.
    \State Matrix B Ordering: Process MatB tiles in column-major order.
    \State Apply algorithmic data repetition (replication factor) patterns.
    \State PLIO formatting: ensure WRD\_LN elements per line.
\EndFor
\end{algorithmic}
\end{algorithm}

\begin{table}[!t]
\centering
\caption{Data Ordering Summary for Stream Generation}
\label{tab:data_ordering}
\setlength{\tabcolsep}{2pt}
\small
\begin{tabular}{|l|c|c|c|}
\hline
\textbf{Data Level} & \textbf{A} & \textbf{B} & \textbf{C} \\
\hline
Elements within sub-tiles & Row-major & Row-major & Row-major \\
\hline
Sub-tiles within tiles & Row-major & Row-major & Row-major \\
\hline
Tiles within blocks & Row-major & Column-major & Column-major \\
\hline
\end{tabular}
\end{table}

Algorithm~\ref{alg:plio_generation} (PLIO\_Stream\_Generation), transforms large input matrices into sequential data streams for the fixed-core AIE-ML graph. Line 1 calculates the number of elements per 128-bit PLIO chunk, WRD\_LN. The number of temporal iterations required to process the full workload, defined by the graph iteration count:

{\footnotesize
\begin{equation}
\label{equ:gic}
\mathrm{GRAPH\_ITER\_CNT} = 
\frac{\mathrm{GEMM\_SIZE\_A} \times \mathrm{GEMM\_SIZE\_B}}
{\mathrm{DIM\_A} \times \mathrm{DIM\_B} \times \mathrm{SPLIT}}.
\end{equation}
}

{\footnotesize
\begin{equation}
\label{equ:rf}
\mathrm{REPLICATION\_FACTOR\_A/B} = 
\frac{\mathrm{GEMM\_SIZE\_B/A}}{\mathrm{DIM\_B/A} \times \mathrm{SPLIT}}.
\end{equation}
}
Here, Matrix A is replicated between tiles after each block processing cycle, while Matrix B is replicated after processing each column within blocks, as illustrated in Figure~\ref{fig:hierarchical_tiling}. The subsequent loop (lines 5-9) processes the matrices using row-major and column-major ordering within this data repetition framework to maximize computational density.

\subsection{Hardware Pipelining and AIE-ML Graph Execution}
This subsection describes the execution of the prepared data streams by the fixed-core AIE-ML graph, detailing the high-speed pipeline protocols that enable overlapping of efficient computation and data transferring.
\begin{figure*}[t]
    \centering
    \includegraphics[width=0.9\textwidth]{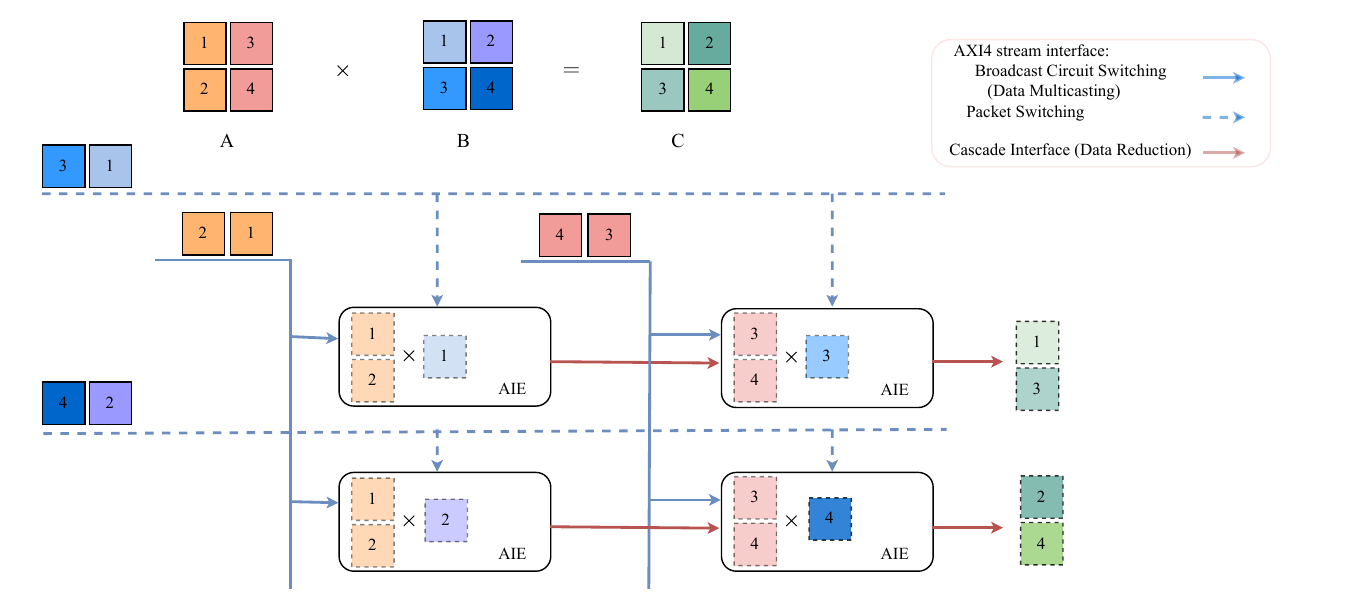}
    \caption{AIE-ML Cores Data Flow for our framework: Fixed AIE-ML Compute Block with Optimized I/O Architecture}
    \label{fig:system_integration}
\end{figure*}

\subsubsection{AIE-ML Engine Graph (Array of AIE-ML Cores)}
Figure~\ref{fig:system_integration} illustrates the execution of the AIE-ML graph that processes the prepared data streams.
The diagram shows the graph's parameterization by CASC\_LN (cascade levels) and SPLIT (parallel splits), with the AIE-ML array connected to the FPGA/PL kernel through PLIO interfaces.
The figure demonstrates how high-speed streaming protocols enable efficient data movement while overlapping computations. The cascade stream chains—visible as horizontal connections between cores—provide 512-bit wide pathways (red arrows) for AIE-to-AIE partial sum reduction, generating c1 and c2 output streams.
It directly implements the dimension reduction from 3D MatMul to 2D array by handling the $\mathrm{GEMM\_SIZE\_AB}$-dimension accumulation through chaining. This minimizes synchronization overhead to achieve Initiation Interval (II) of 1, avoiding the approximately 50\% slower buffer sharing interface.
Input distribution follows specialized routing patterns visible in the figure: Matrix A utilizes broadcast circuit-switching (shown as single source branching to multiple destinations (solid blue arrows)) to simultaneously route a0\_casc* input streams to all SPLIT groups, while Matrix B employs packet switching (depicted as time-multiplexed streams (dashed blue arrows)) to dynamically route different b*\_casc* input streams to different splits, maximizing reuse of the constrained VE2302's PLIO resources.

The AIE-ML graph construction and streaming connections are formalized in Algorithm~\ref{alg:aie}. Lines 2-3 instantiate the core computation graph mmult[SPLIT]. Line 4 creates Matrix A PLIO interfaces, implementing the broadcast circuit-switched distribution. The nested loops (lines 5-16) establish all connections. Matrix A, broadcast to all splits (line 10), enables the data multi-casting. Matrix B split-specific connections in line 11 implement the packet-switched routing. In line 14, output collection gathers results through the cascade streams. Runtime ratios are set in line 8 for performance optimization, completing the graph construction for efficient MatMul execution.

\begin{algorithm}[!t]
\caption{AI Engine Graph Construction}
\label{alg:aie}
\begin{algorithmic}[1]
\State Input: \texttt{CASC\_LN}, \texttt{SPLIT}
\Function{GeMM Constructor}{}
    \State Instantiate mmult[SPLIT]
    \State Create PLIO matA\_inp[CASC\_LN]
    
    \For{$i = 0$ to $SPLIT - 1$}
        \State Create PLIO matB\_inp[CASC\_LN]
        \State Create PLIO matC\_out[i]
        
        \For{$k = 0$ to $CASC\_LN - 1$}
            \State runtime(mmult[i].kernels[k]) $\gets 1.0$
            \State Connect matA\_inp[k] to mmult[i].inA[k]
            \State Connect matB\_inp[idx] to mmult[i].inB[k]
        \EndFor
        
        \State Connect mmult[i].out to matC\_out[i]
    \EndFor
\EndFunction
\end{algorithmic}
\end{algorithm}

\subsubsection{FPGA/PL Kernel (dma\_hls)}

The \texttt{dma\_hls} kernel orchestrates high-speed data transfers between Extrenal DRAM memory and the fixed-core AIE\_ML array via PLIO streams. The kernel is implemented with Vitis HLS, and employs a deadlock-free dataflow design that adheres to PL Resource Conservation by using only lightweight streaming FIFOs, avoiding monolithic buffering, unlike SOTA \cite{CHARM,CHARMtwo}. Algorithm~\ref{alg:pl_execution} details the kernel's high-speed execution. In the algorithm, memory pointers are mapped to NoC DDR4 interfaces through AXI4 Memory-Mapped streams. The top-level DATAFLOW pragma, in Line 4, enables concurrent execution of input/output functions. 
Sequential distribution via modulo addressing in Line 8 enables efficient 128-bit burst data transfers to the AIE-ML array. The design preserves II=1 pipeline efficiency throughout the data path in Line 6.

\begin{algorithm}[!t]
\caption{Top-Level FPGA/PL Execution and Data Flow (DMA\_HLS)}
\label{alg:pl_execution}
\begin{algorithmic}[1]
\Require Memory pointers: matA, matB, matC; Streams: strmInp
\Ensure Streams: strmOut
\Function{DMA\_HLS}{matA, matB, matC, streams}
    \State Constants: NUM\_A\_FILES = 8 
    \State Constants: NUM\_B\_FILES = 16 
    \State \#pragma HLS DATAFLOW
    \For{each memory read transaction}
        \State \#pragma HLS PIPELINE II=1 \Comment{Enforce throughput}
        \State ReadData $\gets$ matX[i]  \Comment{128-bit burst}
        \State stream\_idx $\gets i \bmod$ NUM\_A\_FILES \Comment{Sequential distribution}
        \State Write ReadData to strmOut[stream\_idx]
    \EndFor
    \State out\_C(strmInp\_C, matC)
    \Comment{Deadlock-free Matrix C collection (pairwise writes)}
\EndFunction
\end{algorithmic}
\end{algorithm}

\section{Environmental Setup}

The framework is implemented on the AMD Versal AI Edge VE2302 ACAP (XCVE2302-1LSESFVA784-E), employing AIE-MLv1 architecture with PL components at 312.5 MHz, while limited PLIO resources constrain split (SPLIT) and cascade paths (CASC\_LN). 
A fixed spatial compute block of 16 AIE-ML cores for matrices' workloads of $(32-1024)^3$ with INT16/INT32 precisions are utilized. The 16 cores configuration is used because the area group's 24 registered 128-bit PLIO channels can support it.
The VE2302 ACAP architecture, identified in the compiler by \texttt{\_\_AIE\_ARCH\_\_ == 20}, features native hardware support for \texttt{INT4}, \texttt{INT8}, and \texttt{BFLOAT16} data types, in addition to \texttt{INT16} and \texttt{INT32}. However, our evaluation was constrained by the available software support in the AMD Vitis\texttrademark\ 2024.1 toolchain. Specifically, our design relies on the templated matrix multiplication kernel from the AMD Xilinx DSP Library, which limited the scope of the numerical formats we could report for both simulation and hardware results.

All designs were compiled with AMD Vitis 2024.1. We report full system throughput and resource utilization, which includes the PL and DDR memory controller. This holistic implementation enables a direct comparison with SOTA frameworks that provide on-board results, while excluding simulation-only studies. Finally, power consumption was estimated using the AMD AIE-specific Xilinx Power Estimator (XPE) tool.
\begin{table}[!t]
    \centering
    \caption{Performance and Resource Utilization for $1024^{3}$ INT16 GEMM in \textsc{Tempus}}
    \label{tab:resource_utilization_full}
    \tightesttabular
    \begin{tabular}{p{0.32\linewidth}p{0.23\linewidth}p{0.35\linewidth}}
        \toprule
        \textbf{Metric} & \textbf{Value} & \textbf{Context} \\
        \midrule
        \multicolumn{3}{l}{\textbf{I. Performance and Timing (ms)}} \\
        \midrule
        AIE Cores Used & 16 (47\%) & Fixed, Resource-Invariant \\
        Core Computation ($t_{\text{actual}}$) & 3.537 & Measured execution \\
        Achieved Throughput & 607 GOPS & Derived from latency \\
        Device/XCLBIN Init & 226.928 & One-time setup \\
        Buffer Creation/Mapping & 20.362 & Memory allocation \\
        Kernel/Graph Create & 55.882 & Compilation overhead \\
        Kernel Launch & 0.218 & Launch kernel \\
        PL Tiling & 13.276 & PL Tiling/Replication overhead \\
        Graph Run/DMA Wait & 3.319& Runtime scheduling \\
        Output Sync & 0.013 & Result collection \\
        PL performance & 312.5 & PL performance  \\
        \midrule
        \multicolumn{3}{l}{\textbf{II. Power and Energy}} \\
        \midrule
        AIE Engine Power & 2.381 W & 16 cores active \\
        Memory Power & 3.173 W & B/XRAM + NoC-DDRMC \\
        Total On-Chip Power & 10.677 W & Frugal consumption \\
        Energy Eff. (AIE) & 255 GOPS/W & Core efficiency \\
        Energy Eff. (Total) & 56.87 GOPS/W & System efficiency \\
        \midrule
        \multicolumn{3}{l}{\textbf{III. Resource Utilization}} \\
        \midrule
        LUT & 6.16\% & PL capacity preserved \\
        BRAM & 62.58\% & Streaming FIFOs \\
        URAM & 0.00\% & Resource conservation \\
        DSP & 0.00\% & Resource conservation \\
        CLB Registers & 7.65\% & Resource conservation \\
        \bottomrule
    \end{tabular}
\end{table}
\section{SIMULATION RESULTS AND SUSTAINABILITY ANALYSIS}
This section evaluates the performance and sustainability of the Resource-Invariant Temporal Scaling rectangular GEMM framework.

\subsection{System-Level Characterization: Performance, Power, and Resource Usage}
The operational metrics are detailed in Table~\ref{tab:resource_utilization_full}, representing a system implementation for the $1024^3$ INT16 workload rather than AIE-only simulation. The framework achieves \text{607~GOPS} with a core computation latency of \text{3.537~ms}, and total on-chip power of \text{10.677~W}. A detailed breakdown of the execution timeline reveals that the core computation is highly efficient, underscoring the design's streaming and computation efficiency. Power analysis shows the AIE engines consume only 2.381~W, while memory subsystems (including B/XRAM and NoC-DDRMC) consume \text{3.173~W}, representing a significant portion of the total \text{10.677~W} on-chip power and confirming the I/O-bounded nature of the workload.

\begin{table}[!t]
    \centering
    \caption{Tile Dimension (DIM) Scaling in \textsc{Tempus} for Fixed Workload ($512^3$) in different data types}
    \label{tab:dim_scaling_timing}
    \scriptsize
    \begin{tabular}{cccc}
        \toprule
        \textbf{Type} & \textbf{DIM} & \textbf{Latency (ms)} & \textbf{Throughput (GOPS)} \\
        \midrule
        \multirow{6}{*}{\textbf{INT16}} & 4 & 6.194 & \text{43.338} \\
         & 8 & 3.230 & \text{83.107} \\
         & 16 & 1.811 & \text{148.225} \\
         & 32 & 1.123 & \text{239.034} \\
         & \textbf{64} & 0.792 & \textbf{338.934} \\
         & 128 & 0.586 & \text{458.081} \\
        \midrule
        \multirow{5}{*}{\textbf{INT32}} & 4 & 11.848 & 22.657\\
         & 8 & 6.171 & \text{43.500} \\
         & 16 & 3.225 & \text{83.236} \\
         & 32 & 1.779 & \text{150.891} \\
         & \textbf{64} & 1.150 & \textbf{233.422} \\
        \bottomrule
    \end{tabular}
\end{table}

\subsection{Validation of Temporal Scaling and Workload Analysis}
The performance analysis validates the principle that scalability can be achieved through temporal iteration rather than physical core expansion, demonstrating the efficacy of the resource-invariant approach.

\subsubsection{Tile Dimension Scaling}
Table~\ref{tab:dim_scaling_timing} reveals the critical relationship between the micro-kernel tile size (DIM) and computational efficiency for a $512^3$ workload. Increasing DIM from 4 to 128 improves throughput by $10.5\times$. This shows that providing more local memory per core would directly reduce latency by enabling larger micro-kernels.
Theoretically, DIM could be increased to 256 to further improvement, however the local memory constraint per AIE-ML tile fundamentally caps the practical limit at DIM=128 for INT16. Precision scaling remains predictable: INT32 achieves 233.422\,GOPS at its DIM=64 limit, roughly half the throughput of INT16, reflecting the hardware's $2\times$ data width penalty while confirming \textsc{Tempus}'s robust architectural proficiency within a fixed computational fabric.

\begin{table}[!t]
    \centering
    \caption{Workload scaling in \textsc{Tempus} with Maximum Available DIM in different data types}
    \label{tab:gemm_scaling_timing}
    \scriptsize
    \begin{tabular}{ccccc}
        \toprule
        \textbf{Type} & \textbf{Size} & \textbf{DIM} & \textbf{Latency (ms)} & \textbf{Throughput (GOPS)} \\
        \midrule
        \multirow{7}{*}{INT16} & $32^3$ & 16 & 0.396 & 0.165 \\
         & $64^3$ & 32 & 0.389 & 1.348 \\
         & $128^3$ & 64 & 0.395 & 10.618 \\
         & $256^3$ & 128 & 0.407 & 82.443 \\
         & $512^3$ & 128 & 0.586 & 458.081 \\
         & $768^3$ & 64 & 1.637 & \text{553.433} \\
         & $1024^3$ & 64 & 3.537 & 607.148 \\
        \midrule
        \multirow{7}{*}{INT32} & $32^3$ & 16 & 0.397 & 0.165 \\
         & $64^3$ & 32 & 0.403 & 1.301 \\
         & $128^3$ & 64 & 0.396 & 10.592 \\
         & $256^3$ & 64 & 0.483 & 69.471 \\
         & $512^3$ & 64 & 1.150 & 233.422 \\
          & $768^3$ & 32 & 5.412 & \text{167.400} \\
         & $1024^3$ & 32 & 14.757 & 145.523 \\
        \bottomrule
    \end{tabular}
\end{table}

\subsubsection{Workload Scaling Analysis and Architectural Efficiency}
Table \ref{tab:gemm_scaling_timing} characterizes \textsc{Tempus}'s temporal scaling across an increase in operations $32\,768\times$ (from $32^3$ to $1024^3$). \textsc{Tempus} amortizes fixed overheads, transitioning from sub-optimal efficiency at $32^3$ to a sustained 607 GOPS at $1024^3$ (INT16).
Key insights is that the $512^3$ workload achieves near-ideal scaling at DIM=128, but $1024^3$ is confined to DIM=64, increasing iteration counts and causing a non-linear latency jump to 3.537ms. Notably, precision scaling remains predictable such that INT32 at $1024^3$ is limited to DIM=32, delivering roughly one-quarter of INT16 throughput (145.5 GOPS) due to the $2\times$ data width penalty. This adaptive scaling proves \textsc{Tempus} automatically respects hardware boundaries, delivering predictable, sustainable performance for real-time edge AI.

\begin{table}[!t]
    \centering
    \caption{PL Resource Utilization and Power Consistency of \textsc{Tempus} for INT16, URAM/DSP utilization is 0.00\% across all workloads}
    \label{tab:consistency}
    \scriptsize
    \setlength{\tabcolsep}{2pt}
    \begin{tabular}{@{}lccccc@{}}
        \toprule
        \textbf{Workload} & \textbf{Total On-Chip Power(W)} & \textbf{LUT(\%)} & \textbf{BRAM(\%)} & \textbf{CLB Regs(\%)}\\
        \midrule
        $32^3$ & 10.698 & 6.09 & 62.58 & 7.63\\
        $64^3$ & 10.639  & 6.11 & 62.58 & 7.64\\
        $128^3$ & 10.315  & 6.13 & 62.58 & 7.65\\
        $256^3$ & 10.692  & 6.11 & 62.58 & 7.64\\
        $512^3$ & 10.661  & 6.18 & 62.58  & 7.65\\
        $768^3$ & 10.631  & 6.20 & 62.58  & 7.67\\
        $1024^3$ & 10.677  & 6.16 & 62.58  & 7.65\\
        \midrule
        $8\times32\times8$ & 10.701  & 6.11 & 62.58  & 7.64\\
        $128\times768\times64$ & 10.236  & 6.14 & 62.58  & 7.66\\
        $512\times64\times512$ & 10.281  & 6.11 & 62.58  & 7.64\\
        $512\times1024\times512$ & 10.680  & 6.17 & 62.58  & 7.65\\
        $128\times768\times3072$ & 10.721  & 6.17 & 62.58  & 7.66\\
        $768\times3072\times768$ & 10.788  & 6.18 & 62.58  & 7.68\\

        $8\times1024\times1024$& 10.282  & 6.15 & 62.58  & 7.65\\
        $8\times2048\times2048$& 10.703  & 6.15 & 62.58  & 7.65\\
        $8\times4096\times4096$& 10.715 & 6.19 & 62.58  & 7.65\\

        \bottomrule
    \end{tabular}
\end{table}

\begin{table*}[!t]
\centering
\scriptsize
\caption{Comprehensive Comparative Analysis of Throughput, Power, and Platform-Aware Utility for $1024^{3}$ INT16 GEMM \cite{CHARMtwo,ARIES,AutoMM,CHARM,AutoSA}}
\label{tab:comparison_combined}
\begin{tabular}{lcccccccccccc}
\toprule
\textbf{Framework} & \textbf{Cores} & \textbf{Lat.(ms)} & \textbf{TOPS} & \textbf{Pwr(W)} & \textbf{U\%(1)} & \textbf{PLIO} & \textbf{T/C(2)} & \textbf{T/P(3)} & \textbf{C-Fru(4)} & \textbf{P-Fru(5)} & \textbf{I-Fru(6)} & \textbf{PAU(n)(7)} \\
\midrule
\textbf{\textsc{Tempus} (Temporal)} & \textbf{16} & 3.537 & 0.607 & \textbf{10.677} & \textbf{0.00} & \textbf{26} & \text{0.038} & \text{0.057} & \textbf{22.0$\times$} & \textbf{7.1$\times$} & \textbf{6.3$\times$} & \textbf{211.2$\times$} \\
\textsc{ARIES} (Spatial) & 352 & 0.1354 & 15.86 & 76.30 & 76.03 & 164 & 0.045 & 0.208 & 1.0$\times$ & 1.0$\times$ & 1.0$\times$ & \text{1.0} \\
\textsc{CHARM} 2.0 (Spatial) & 288 & 0.2141 & 10.03 & 64.80 & 82.94 & 120 & 0.035 & 0.155 & 1.2$\times$ & 1.1$\times$ & 1.4$\times$ & \text{1.2$\times$} \\
\textsc{AutoMM} (Spatial) & 288 & 0.2859 & 7.51 & 56.80 & 82.94 & 120 & 0.026 & 0.132 & 1.2$\times$ & 1.3$\times$ & 1.4$\times$ & \text{1.1$\times$} \\
\textsc{AutoSA} (Spatial) & -- & 0.6298 & 3.41 & 84.90 & -- & -- & -- & 
        0.0401 & -- & -- & -- & --\\
\bottomrule
\end{tabular}
\vspace{2pt}

{\footnotesize \textit{Notes:  (1) URAM\% Utilization. (2) TOPS/Core density. (3) TOPS/Power efficiency (AI Efficiency). (4) C-Fru: Core Frugality. (5) P-Fru: Power Frugality, (6) I-Fru: I/O Frugality (PLIO). (7) Platform-Aware Utility Factor.}}
\end{table*}

\begin{table*}[!t]
\centering
\scriptsize
\caption{Compute, Resource \& Power Strengths Comparison}
\label{tab:versal-strengths}
\begin{tabular}{p{3cm}p{2.6cm}p{3.1cm}p{3.6cm}}
\toprule
\textbf{Feature} & \textbf{VCK190 (VC1902)} & \textbf{VE2802} & \textbf{VE2302} \\
\midrule
\multicolumn{4}{l}{\textbf{AI Compute \& Efficiency}} \\
\midrule
AI Engine Type & 1st Gen AI Engine\cite{AMD_UG1079_2025} & AIE-ML v2\cite{AMD_Versal_AI_Edge_PSG_2024} & AIE-ML\cite{Xilinx_Versal_AI_Edge_WP518_2021} \\
Cores & 400 & 304 & 34 \\
Peak AIE INT16 Performance & 64 TOPS & 101 TOPS & 11.5 TOPS \\ 
AI Efficiency (TOPS/W) & 0.71--1.28 & 2.69 & 1.15--1.53 \\
\midrule
\multicolumn{4}{l}{\textbf{Power \& Thermal}} \\
\midrule
Total Chip Power (TCP) & 100--180 W & Up to 75 W & 15--20 W \\
\midrule
\multicolumn{4}{l}{\textbf{Programmable Logic \& Resources}} \\
\midrule
System Logic Cells & 1,968K & 1,139K & 328K \\
DSP Engines & 1,968 & 1,312 & 464 \\
\midrule
\multicolumn{4}{l}{\textbf{External Memory Support}} \\
\midrule
DDR4 Support & 8 GB @ 3200 Mb/s & Up to 16 GB & 4 GB (64-bit, upgradable to 8 GB) \\
LPDDR4 Support & 8 GB @ 3900 Mb/s & 12 GB @ 3733 Mb/s (192-bit) & 4 GB (64-bit) \\

\bottomrule
\end{tabular}
\vspace{4pt}

\parbox{\linewidth}{\footnotesize General Notes: INT16 performance inferred from INT8 specifications ($\frac{1}{2} \times$ INT8). AI Efficiency = Peak INT8 TOPS $\div$ Max TCP.}
\end{table*}

\subsection{Resource and Power Invariance}
Table~\ref{tab:consistency} demonstrates that \textsc{Tempus} maintains strict resource invariance across exponential workload growth. Total on-chip power stays frugal at $\sim$10.6\,W, and critical PL resources (DSP, URAM) remain at 0.00\% utilization. This contrasts sharply with SOTA spatial designs, which saturate resources (e.g., \textsc{CHARM} 2.0 uses 82.94\% of URAM). Low LUT and BRAM usage—due only to lightweight streaming FIFOs (FIFO\_depth=16, outstanding=32, BURST=32), preserves PL capacity for heterogeneous orchestration of essential kernels like Softmax and LayerNorm in complete model pipelines. Further reductions in FIFO depth, outstanding transactions, and burst size could yield even greater BRAM savings.

\section{Comparative Sustainability and Resource Frugality}

We evaluate the resource-invariant \textsc{Tempus} framework against spatial SOTA frameworks (\textsc{ARIES}, \textsc{CHARM} 2.0, \textsc{AutoMM}, \textsc{AutoSA}) in Table~\ref{tab:comparison_combined}. These baselines prioritize peak throughput on high-end Versal devices (VCK190/VE2802, $\sim$300$-$400 cores), which differ from \textsc{Tempus}'s VE2302 in compute efficiency, and memory capacity (see Table~\ref{tab:versal-strengths}). However, they fail in edge-class SoCs because reducing core counts violates their assumptions about spatial parallelism, leading to compilation failure. To enable fair comparison despite hardware asymmetry, we introduce platform-aware metrics that decouple algorithmic efficiency from absolute resource budgets.

\subsection{Platform-Aware Utility (PAU(n))}
Architectural proficiency is evaluated by measuring the extracted computational work  relative to the total physical potential and resource footprint of the deployment platform (cores, power, I/O, and peak throughput). It rewards designs that perform well on resource-constrained boards and penalizes brute-force spatial arrays. PAU is defined as:

{\footnotesize
\[
PAU = \frac{\text{TOPS}}{\text{Cores} \times \text{Power (W)} \times \text{PLIO} \times \text{Theoretical Peak (Pk)}}
\]}

To highlight \textsc{Tempus}'s architectural advantage, we define the \text{Platform-Aware Utility Factor} $n = PAU_{\text{other}} / PAU_{\textsc{ARIES}}$, where $n > 1$ indicates higher utility than the SOTA \textsc{ARIES} baseline. Table \ref{tab:comparison_combined} shows that although \textsc{Tempus} has higher absolute latency ($3.537\,\text{ms}$ on VE2302 vs. $0.135\,\text{ms}$ on VCK190 for \textsc{ARIES}), it achieves a $211.2\times$ higher utility factor, avoiding the utilization collapse inherent in rigid spatial architectures.

\subsection{Resource-Invariant Frugality and Heterogeneous Orchestration}
The sustainable execution of \textsc{Tempus} is characterized through its multi-dimensional frugality across core, power, and I/O domains. 
These metrics are defined as:

{\footnotesize
\[
\mathcal{C}\text{-Fru} = \frac{\text{Cores}_{\text{other}}}{\text{Cores}_{\text{\textsc{Tempus}}}},\quad
\mathcal{P}\text{-Fru} = \frac{\text{Power}_{\text{other}}}{\text{Power}_{\text{\textsc{Tempus}}}}
\]}
{\footnotesize
\[
\mathcal{I}\text{-Fru} = \frac{\text{PLIO}_{\text{other}}}{\text{PLIO}_{\text{\textsc{Tempus}}}}
\]}

As detailed in Table~\ref{tab:comparison_combined}, \textsc{Tempus} achieves $22.0\times$ Core Frugality, $7.1\times$ Power Frugality, and $6.3\times$ I/O Frugality relative to the \textsc{ARIES} baseline.
While spatial scaling SOTA frameworks reach an architectural dead end on compact devices by saturating $76\%$--$83\%$ of on-chip URAM for a single GEMM kernel, \textsc{Tempus} maintains \text{0.00\% URAM/DSP utilization} components. 
The I/O advantage is equally significant and invariant. Our hybrid streaming approach—combining packet switching for dynamic time-multiplexing of data streams with cascade streaming that eliminates redundant I/O for intermediate results—ensures that I/O constraints never limit scalability.
This resource invariance is essential for heterogeneous orchestration, as it preserves the programmable logic fabric and memory resources for concurrent integration of other critical model kernels (e.g., Softmax, Layer Normalization).

\subsection{Normalized Computational Efficiencies ($T/\mathcal{C}$, $T/\mathcal{P}$):}
Throughput per core ($T/\mathcal{C}$) and performance per watt ($T/\mathcal{P}$) are reported to characterize computational density and AI efficiency within strict thermal boundaries:

{\footnotesize
\[
T/\mathcal{C} = \frac{\text{TOPS}}{\text{Cores}}, \qquad
T/\mathcal{P} = \frac{\text{TOPS}}{\text{Power (W)}}
\]}

Despite operating with a hardware budget nearly $6\times$ lower than the spatial reference (11.5 vs. 64.0 Peak TOPS), \textsc{Tempus} achieves competitive $T/\mathcal{P}$ and maintains high computational density. This efficiency ensures sustainable, energy-sensitive execution for foundation models where traditional massive parallelism would lead to thermal failure on edge devices.

\begin{table*}[!t]
\centering
\caption{Shape-Agnostic \textsc{TEMPUS} Performance: Rectangular GEMMs and Timing-Equivalent Cubic Workloads (INT16)}
\label{tab:rectangular_cubic_equivalence}
\scriptsize
\begin{tabular}{lcccc}
\toprule
\textbf{Architectural Role} & \textbf{Rectangular GEMM} & \textbf{Rectangular Latency (ms)} & \textbf{Cubic Equivalent} & \textbf{Cubic Latency (ms)} \\
\midrule
\multicolumn{5}{c}{\textbf{Decoding Phase (Attention Projection Layers)} (narrow shapes)} \\
\midrule
Small/Mobile LLM (e.g., Pythia/MobileLLM) & $8\times1024\times1024\ (DIM=128)$ & 0.604 & $192^3 (DIM=96)$ & 0.403\\
Small/Mobile LLM (TinyLlama/Gemma) & $8\times2048\times2048\ (DIM=64)$ & 1.527 & $768^3(DIM=64)$ &  1.637\\

Production LLM (LLaMA-2 7B)        & $8\times4096\times4096\ (DIM=32)$ & 5.241 & $1024^3(DIM=64)$ & 3.537 \\
\midrule
\multicolumn{5}{c}{\textbf{Attention Head Logic} (fragmented shapes)} \\
\midrule
Tiny head (experimental)           & $8\times32\times8\ (DIM=4)$        & 0.394 & $32^3(DIM=4)$  & 0.396 \\
BERT-base single head \cite{devlin2018bert} & $128\times768\times64\ (DIM=64)$   &  0.394 & $192^3(DIM=96)$ &  0.403\\
Attention score matrix (seq=512) \cite{vaswani2017attention}   & $512\times64\times512\ (DIM=128)$    &  0.446& $256^3(DIM=128)$ & 0.407 \\
Vision Transformer (ViT) head      & $128\times128\times128\ (DIM=64)$   & 0.395 & $128^3(DIM=64)$ & 0.395 \\
\midrule
\multicolumn{5}{c}{\textbf{Feed‑forward networks (FFN)} (wide shapes)} \\
\midrule
BERT-base FFN Up-projection \cite{devlin2018bert} & $128\times768\times3072\ (DIM=96)$ & 1.258 & $768^3(DIM=64)$ & 1.637 \\
Production-scale mid-size \cite{koroteev2021bert} & $512\times1024\times512\ (DIM=64)$ & 1.147 & $512^3(DIM=128)$ &  0.586\\
BERT-base FFN expansion \cite{devlin2018bert}    & $768\times3072\times768\ (DIM=16)$ & 19.674 & $1216^3(DIM=32)$ & 9.907  \\

\bottomrule
\end{tabular}
\vspace{2pt}

\end{table*}

\subsection{Shape-Agnostic Efficiency Across LLM Architectures}
To demonstrate generality, we evaluate \textsc{Tempus} on representative rectangular GEMM shapes from LLM components: decoding projection layers, multi‑head attention, and feed‑forward networks (FFNs). Table~\ref{tab:rectangular_cubic_equivalence} compares each rectangular workload against a cubic shape whose \text{latency matches} the rectangular one. The results show that \textsc{Tempus} maintains predictable, high efficiency across unbalanced dimensions where spatial frameworks suffer “utilization collapse” and performance drops up to $5760\times$.

\paragraph{Decoding phase (narrow shapes)}
In LLMs, while training utilizes massive batches of 2 million tokens \cite{zhang2024tinyllama} to saturate hardware, mobile deployment requires efficiency at $GEMM\_SIZE\_A \leq 8$. Therefore, the decoding phase models a single-token inference step where the input is a single vector \cite{dao2023flashattention}. For Mobile LLMs like TinyLlama-1.1B \cite{zhang2024tinyllama}, and \text{Pythia-410M}, and Gemma-2B \cite{team2024gemma}, the hidden dimension is 1024, and 2048. Similarly, the larger LLaMA-2 7B has a ($8 \times 4096 \times 4096$) projection. As tabulated, the small difference between rectangular and cubic comes from a reduction in micro‑kernel tile size (DIM) from the optimal DIM=128 to DIM=64, forced by the local memory limit, not a fundamental inefficiency.

\paragraph{Attention heads (fragmented shapes)}
Multi-head attention splits embeddings into smaller, fragmented heads \cite{vaswani2017attention}. This creates rectangular GEMMs ranging from tiny experimental shapes ($8 \times 32 \times 8$) to standard BERT-base single heads ($128 \times 768 \times 64$). Fragmented heads nearly identical to the cubic equivalent. This proves \textsc{Tempus} is resilient to narrow dimensions that cause massive spatial arrays to underutilize resources.

\paragraph{Feed‑forward networks (wide shapes)}
The feed-forward network (FFN) creates "wide" or "tall" rectangular GEMMs during up-projections and expansions. For BERT-Base, the up-projection of $128 \times 768 \times 3072$ and the full expansion of $768 \times 3072 \times 768$ represent production-scale workloads \cite{devlin2018bert, koroteev2021bert}. These results confirm that \textsc{Tempus} is shape‑agnostic: it extracts high utility from rectangular LLM workloads without the architectural mismatches of spatial accelerators.

\section{CONCLUSION}
This work introduces \textsc{\textsc{Tempus}}, the first resource-invariant GEMM streaming framework for the Versal AI Edge VE2302. Mapping 3D MatMul onto a fixed 2D array, it achieves 607 GOPS using only 16 AIE-ML cores via algorithmic iteration. Versus spatial SOTA (\textsc{ARIES}), \textsc{\textsc{Tempus}} delivers a $211.2\times$ higher PAU prominence factor, formally justifying its architectural superiority for edge deployment. The framework ensures sustainable performance through multi-dimensional frugality ($22.0\times$ core, $7.1\times$ power, $6.3\times$ I/O). Critically, resource utilization remains invariant across workloads at $10.677$ W. This conservation preserves programmable logic for heterogeneous orchestration of non-GEMM kernels (Softmax, Layer-normalization) required for foundation model inference.

\bibliographystyle{IEEEtran}
\bibliography{ref.bib}

\end{document}